\documentclass[copyright,creativecommons]{eptcs}
\providecommand{\event}{F-IDE 2021} 
\usepackage{underscore}           

\usepackage{why3lang}

\usepackage{tikz}
\usetikzlibrary{matrix,positioning,fit}

\newcommand{\limpl}{\rightarrow}
\newcommand{\result}{\textsf{result}}
\newcommand{\WP}{\textrm{WP}}
\newcommand\impwhilenovar[4]{\texttt{while}~#1~\texttt{do}~%
  \texttt{invariant}~\texttt{\{}~#2~\texttt{\}}~%
  \texttt{writes}~\texttt{\{}~#3~\texttt{\}}~%
  #4~\texttt{done}}

\title{Explaining Counterexamples\\
  with Giant-Step Assertion Checking%
  \thanks{This work was partly funded by collaborations between Inria and the companies AdaCore
    (Paris, France) and Mitsubishi Electric R\&D Centre Europe (Rennes, France).}}

\author{Benedikt Becker \and Cl\'audio Belo Louren\c{c}o \and Claude March\'e
  \institute{Universit\'e Paris-Saclay, CNRS, Inria, LMF, 91405, Orsay, France}
}

\def\titlerunning{Explaining Counterexamples}
\def\authorrunning{Becker, Belo Louren\c{c}o \& March\'e}

\begin{document}
\maketitle

\begin{abstract}
  Identifying the cause of a proof failure during deductive verification of
  programs is hard: it may be due to an incorrectness in the program, an
  incompleteness in the program annotations, or an incompleteness of the prover.
  The changes needed to resolve a proof failure depend on its category, but the
  prover cannot provide any help on the categorisation.
  When using an SMT solver to discharge a proof obligation, that solver can
  propose a model from a failed attempt, from which a possible counterexample
  can be derived. But the counterexample may be invalid, in which case it may
  add more confusion than help.
  To check the validity of a counterexample and to categorise the proof failure,
  we propose the comparison between the run-time assertion-checking (RAC)
  executions under two different semantics, using the counterexample as an
  oracle. The first RAC execution follows the normal program semantics, and a
  violation of a program annotation indicates an incorrectness in the program. The
  second RAC execution follows a novel ``giant-step'' semantics that does not
  execute loops nor function calls but instead retrieves return values and values of
  modified variables from the oracle. A violation of the program annotations
  only observed under giant-step execution characterises an incompleteness of
  the program annotations.
  We implemented this approach in the Why3 platform for deductive program
  verification and evaluated it using examples from prior literature.
\end{abstract}

\section{Introduction}

Deductive program verification aims at checking that a given program respects a
given functional behaviour. The expected behaviour is expressed formally by
logical assertions, principally pre-conditions and post-conditions on procedures
and functions, and invariants on loops.  The verification process consists in
generating, from the code and the formal annotations, a set of logic formulas
called \emph{verification conditions} (VCs), typically via a \emph{Weakest
  Precondition Calculus} (WP)~\cite{Dijkstra76}. If the VCs are
proved valid, then the program is guaranteed to satisfy its specification.
Deductive verification environments like Dafny~\cite{leino14fide},
OpenJML~\cite{cok14}, or Why3~\cite{bobot14sttt}, pass the VCs to automatic
theorem provers usually based on \emph{Satisfiability Modulo Theories} (SMT),
such as Alt-Ergo~\cite{alt-ergo}, CVC4~\cite{barrett11cade} and
Z3~\cite{demoura08tacas}.  Due to the nature of these solvers, the conclusion of
each VC is negated to form a proof goal and the background solver is queried for
satisfiability.
Since these solvers are assumed to be sound when the answer is ``unsat'', one can conclude
that the VC is valid when that is indeed the answer.

\begin{figure}[tb]
  \centering
    \tikzstyle{data} = [rectangle, rounded corners, draw, fill=black!5]
  \begin{tikzpicture}
    \node[data](program) at (0,2) {Program};
    \node[data](vcs) at (4,2) {VCs};
    \node[data](smtinput) at (8,2) {SMT input};
    \node[data](categ) at (0,0) {Categorisation};
    \node[data](ce) at (4,0) {Candidate CE};
    \node[data](model) at (8,0) {Prover model};
    \node(smtsolver) at (10.75,1) {SMT solver};
    \node(proved) at (13,0) {VC proved};
    \path[-latex,every node/.append style={font=\scriptsize}]
      (program) edge node[above] {VC} node[below] {generator} (vcs)
      (vcs) edge node[above] {transformations} node[below] {(goal negated)} (smtinput)
      (smtinput) edge[out=east,in=north,looseness=1] (smtsolver)
      (smtsolver) edge[out=south,in=west,looseness=1] node[above right]{unsat} (proved)
      (smtsolver) edge[out=south,in=east,looseness=1] node[above left]{other} (model)
      (model) edge node[above] {CE} node[below]
        {generation~\cite{dailler18jlamp}} (ce)
      (ce) edge node[above] {this} node[below] {work} (categ)
      ;
  \end{tikzpicture}
  \caption{Pipeline of the counterexample generation and categorisation}
  \label{fig:ce-gen}
\end{figure}
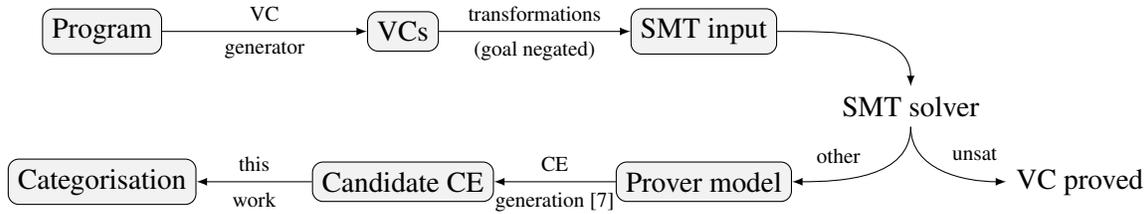

In this paper we address the case when the solver does not answer ``unsat'',
and provide a method to explain why the proof could not be completed. The solver
may give instead several other answers: at best it answers ``sat'', possibly
with a \emph{model}, which is a collection of values for the variables in the
goal. As displayed in Fig.~\ref{fig:ce-gen}, we rely on an approach by Dailler
et al~\cite{dailler18jlamp} to turn such a model into a \emph{candidate counterexample},
which is essentially a collection of triples (variable, source location, value)
representing the different values taken by a program variable during an
execution. Such a counterexample may indicate two different problems:
(i) a \emph{non-conformance}, where the code does not satisfy one of its annotations;
(ii) a \emph{subcontract weakness}, where the annotations are not appropriate
  to achieve a proof (typically a weakness of a loop invariant or
  post-condition).
Unfortunately there is no direct way to distinguish these cases.  Other answers
of the prover are even less informative: the answer ``unknown'' replaces ``sat''
when the solver is incomplete, for example in presence of quantified hypotheses
or formulas involving non-linear arithmetic, or the solver reaches a time limit
or a memory limit. In these cases, the solver may as well propose a model, but
without any guarantee about its validity. Summing up, for any other answer than
``unsat'', there is a need to validate the resulting candidate counterexample
and categorise it as non-conformance or subcontract weakness.

We propose the categorisation of counterexamples using a novel
notion of assertion checking, called \emph{giant-step} assertion checking.
Let us illustrate the idea on the following toy program, that operates on a
global variable~\w{x}.
\begin{why3num}
let set_x (n:int) : unit ensures { x > n }
  x <- n + 1

let main () : unit
  x <- 0; set_x 2; assert { x = 3 }
\end{why3num}
The VC for the function \w{main} is
$\forall x.~ x = 0 \limpl \forall x'.~ x' > 2 \limpl x' = 3$ where $x'$ denotes
the new value of \w{x} after the call to \w{set_x}, and the premise of the
second implication comes from the post-condition of \w{set_x}. The query sent to
the solver is $x = 0 \land x' > 2 \land \lnot (x' = 3)$, from which the solver
would typically answer ``sat'' with the model $\{x=0,x' = 4\}$, corresponding to
a counterexample where \w{x} is 0 after the assignment and 4 after the call to
\w{set_x}. If we proceed to a regular assertion-checking execution of \w{main},
no issue is reported: both the post-condition of \w{set_x} and the assertion in
\w{main} are valid. Our giant-step assertion checking executes calls to
sub-programs in a single step, selecting the values of modified variables from
the candidate counterexample: it handles
the call \w{set_x 2} by extracting the new value for \w{x} from the
counterexample, here $4$, and checking the post-condition, which is correct
here.  The execution then fails because the assertion is wrong. Since the standard
assertion checking is OK but the giant-step one fails, we report a subcontract
weakness. This is the expected categorisation, suggesting the user to improve
the contract of \w{set_x}, in this case by stating a more precise
post-condition.  Giant-step assertion checking also executes loops by a single
giant step to identify subcontract weaknesses in loop invariants.

In Section~\ref{sec:gs}, we explain in more details the concept of giant-step
execution, and how we use it to categorise counterexamples.  In
Section~\ref{sec:exp}, we present our implementation, and experimental results,
that were conducted within the Why3 environment~\cite{bobot14sttt}.  We discuss
related work and future work in Section~\ref{sec:concl}. The technical details
that we skip here due to lack of space are available in a research
report~\cite{becker21rr}.

\section{Giant-Step Assertion Checking}
\label{sec:gs}

We consider here a basic programming language with mutable variables, function calls
and while loops, where functions are annotated with contracts (a pre-condition and a
post-condition) and loops are annotated with a loop invariant. The language also includes
a standard \w{assert} statement, as shown in the example above.

\paragraph{Giant-step assertion checking.} It corresponds to standard runtime
assertion checking (RAC), in the sense that annotations (i.e., assertions,
pre- and post-conditions, and loop invariants) are checked when they are
encountered during execution. It differs from the standard RAC in the treatment
of function calls and loops: instead of executing their body statements, they
only involve a single, ``giant'' step. Giant-step execution is parameterised by
an \emph{oracle}, from which one retrieves the values of variables that are
written by a function call or a loop.  These written variables could be
automatically inferred, but for simplicity we require them to be indicated by a
\w{writes} clause. Therefore, a function declaration has the form:

\begin{why3}
let f ($x_1$: $\tau_1$) $\cdots$ ($x_n$: $\tau_n$) : $\tau$
  requires { $\phi_{pre}$ } ensures { $\phi_{post}$ } writes { $y_1,\ldots,y_k$ } = $e$
\end{why3}

\noindent and a loop has the form:

\begin{why3}
while $e_1$ invariant { $\phi_{inv}$ } writes { $y_1,\ldots,y_k$ } do $e_2$ done
\end{why3}

\begin{figure}[t]
\begin{lstlisting}[numbers=left,numberstyle=\footnotesize,
  basicstyle={\upshape\sffamily},mathescape=true,columns=fullflexible,literate={{-}{-}{1}}]
evaluate(context,oracle,$e$):
 if expression $e$ is:
 $\bullet$ a function call $(\texttt{f}~e_1~\cdots~e_n)$:
     query the context for declaration of $f$:
       parameters $x_1 \cdots x_n$, pre-condition $\phi_{pre}$, writes $y_1 \cdots y_k$ post-condition $\phi_{post}$
     evaluate each $e_i$ to value $t_i$, and set each $x_i$ to $t_i$ in the context
     evaluate $\phi_{pre}$, if false report assertion failure
     query the oracle for values $v_1 \cdots v_k$ for $y_1 \cdots y_k$ and $v$ for $\result$, assign them in the context
     evaluate $\phi_{post}$, if false report execution stuck
     return value $v$
 $\bullet$  a loop $(\texttt{while}~c~\texttt{invariant \{}~\phi_{inv}~\texttt{\} writes \{}~y_1\cdots y_k~\texttt{\} do}~e~\texttt{done})$ :
     evaluate $\phi_{inv}$, if false report assertion failure
     query the oracle for values $v_1 \cdots v_k$ for $y_1 \cdots y_k$, assign in the context
     evaluate $\phi_{inv}$, if false report execution stuck
     evaluate $c$, if false return
     evaluate loop body $e$
     evaluate $\phi_{inv}$, if false report assertion failure
     report execution stuck
 $\bullet$ otherwise: as standard assertion-checking execution
\end{lstlisting}
\caption{Pseudo-code for giant-step assertion-checking execution}
\label{fig:exec}
\end{figure}

\begin{figure}
\[
\begin{array}{l}
  \WP(\impwhilenovar{e_1}{\phi_{inv}}{y_1,\ldots,y_k}{e_2},Q) = \\
  \quad \phi_{inv} ~\land \forall v_1,\ldots,v_k.~
  (\phi_{inv} \limpl \WP(e_1,(\result\limpl\WP(e_2,\phi_{inv}))\land(\lnot\result\limpl Q)))[y_j\leftarrow v_j] \\
\\
  \WP((f~t_1~\cdots~t_n),Q) = \phi_{pre}[x_i\leftarrow t_i] ~\land \forall v_1,\ldots,v_k,v.~ (\phi_{post}[x_i\leftarrow t_i] \limpl Q)[y_j\leftarrow v_j][\result\leftarrow v]
\end{array}
\]
  where $v,v_1,\ldots,v_k$ are fresh variables.
  \caption{Rules for WP  computation, loops and function calls}
  \label{fig:wploopscalls}
\end{figure}

Figure~\ref{fig:exec} presents a pseudo-code for giant-step evaluation of a
program expression $e$ (see~\cite{becker21rr} for a formal presentation). This
execution form is inspired by the weakest precondition calculus, specifically
the rules for calls and loops that we remind in
Figure~\ref{fig:wploopscalls}. The execution may fail or get stuck in a number
of situations:
\begin{itemize}
\item Line 7: if the pre-condition of a call to $f$ is false, the
  execution must fail (as in standard RAC).

\item Line 9: if the post-condition of a call to $f$ is false,
  the values from the oracle are incompatible with the postcondition
 of $f$. A stuck execution is reported, and the counterexample will be declared invalid.

\item Line 12: as in standard assertion checking, a failure is reported for the
invariant initialisation.

\item Line 14: if the invariant is false, the oracle does not
provide valid values to continue the execution. A stuck execution is reported, and the
counterexample will be invalid.

\item Line 15: if the loop condition is false after setting the values of
written variables in the context, the oracle covers an
execution that goes beyond the loop, so we just terminate its execution.

\item Line 17: if invariant is not preserved, we report an assertion failure.

\item Line 18: if invariant is preserved, it means the oracle is not
appropriate for identifying any failure, we report a stuck execution.
\end{itemize}

\paragraph{Categorisation of counterexample.}
As shown on Fig.~\ref{fig:ce-gen}, assume that a VC for a program is not
validated by some SMT solver, which returns a model, which is turned into a
candidate counterexample. We categorise this counterexample
as follows (stopping when the first statement is met):
\begin{enumerate}
\item Run the standard RAC on the enclosing function of the VC, with arguments and values of global variables taken from the counterexample. If the result is
  \begin{enumerate}
  \item a failure at the same source location as the VC, we report a
    non-conformance: the code does not satisfy the corresponding
    annotation.
  \item a failure at a different source location as the VC, the counterexample
    is bad (is not suitable to explain the failed proof), although it deserves
    to be reported as a non-conformance elsewhere in the code: it exposes an
    execution where the program does not satisfy some annotation.
  \end{enumerate}
\item Run the giant-step RAC on the enclosing function of the VC, with inputs
  and written variables given by the counterexample seen as an oracle. If the result is
  \begin{enumerate}
  \item a failure, we report a sub-contract weakness: some
    post-condition or loop invariant in the context is too weak, as in the introductory toy
    example.
  \item a stuck execution, the counterexample is invalid and discarded: one of the premises of the
    VC is not satisfied, and we can suspect a prover incompleteness.
  \item a normal execution, the counterexample is discarded: it satisfies the
    conclusion of the VC.
  \end{enumerate}
\end{enumerate}

\section{Experiments}
\label{sec:exp}

\begin{figure}[tb]
  \begin{lstlisting}[language=why3,numbers=left,mathescape]
  let isqrt (n: int)
    requires { 0 <= n <= 10000 }
    ensures { result * result <= n < (result + 1) * (result + 1) }
  = let ref r = n in
    let ref y = n * n in
    let ref z = -2 * n + 1 in
    while y > n do
      invariant $I_1$ { 0 <= r <= n }
      invariant $I_2$ { y = r * r }
      invariant $I_3$ { n < (r+1) * (r+1) }
      invariant $I_4$ { z = -2 * r + 1 }
      y <- y + z; z <- z + 2; r <- r - 1
    done;
    r\end{lstlisting}
  \caption{Computation of the integer square root, adapted from C code of Petiot~\cite{petiot18fac}}
  \label{fig:example-isqrt}
\end{figure}

We implemented our approach in the Why3 platform for deductive program
verification,
and tested it by reproducing the experiments from prior literature
about the categorisation of proof failures (Petiot et al.~\cite{petiot18fac}).
These experiments covered three example programs written in C
with ACSL program annotations, and the verification was carried out in the
Frama-C framework. The experiments comprised modifications to the programs that
introduced proof failures, which were then categorised using their approach. We
translated the C programs to WhyML and were able to reproduce the
categorisations with our approach for all 16 modifications that were applicable
to the WhyML program.

The experiments by Dailler et
al.~\cite{dailler18jlamp} were written in Ada/SPARK, which uses Why3 for
deductive verification. Their ``Riposte'' testsuite contains 24 programs with
247 checks. The integration of our approach with SPARK is work in
progress. Currently, we are able to identify in this testsuite 14 wrong
counterexamples, 57 non-conformities, and 22 other cases classified as
``either non-conformity or subcontract-weaknesses'' (an imprecise answer which
results when the standard RAC is non-conclusive~\cite{becker21rr}). The counterexamples of 154
checks could not yet be categorised.

Let us illustrate our experiments on one of Petiot's
examples~\cite{petiot18fac} in C, a function that calculates the integer
square root. Our translation of the C program to WhyML is shown in
Figure~\ref{fig:example-isqrt}. The result for parameter $n$ is an integer $r$
such that $r^2 \le n < (r+1)^2$. The variable $r$ is initialised by $n$ as an
over-approximation of the result, the variable $y$ by $n^2$, and $z$ by $-2*n +
1$. During execution of the while loop, the value of $r$ is decremented and the
value of $y$ is kept at $r^2$, while maintaining $n < (r+1)^2$. When the loop
condition becomes false, $r$ contains the largest integer such that $r^2 \le n$.
The VC for function \w{isqrt} is split by Why3 into nine
verification goals: two for the initialisation and preservation of each of
the four loop invariant, and one for the post-condition. The validity of
the program is proven in Why3 (we used the Z3 solver to prove the goals and
generate counterexamples).

We reproduce here two variations of that program, that introduce proof failures.
First,
changing the first assignment in line~12 to \w{y <- y - z} leads to a proof failure
for the preservation of invariant~$I_2$. The counterexample gives the value~4
for the argument \w{n} of function \w{isqrt}. The standard RAC execution fails
after the first loop iteration when checking the preservation of invariant
$I_2$, and the proof failure is categorised as non-conformity.
Second, removing loop invariant $I_3$ leads to a proof failure for the
postcondition.  The counterexample gives the value~1 for the argument \w{n}. The
standard RAC execution terminates normally with the correct result of~1. The
giant-step RAC execution initialises the variables \w{r}, \w{y}, and \w{z} with
values~1, 1, and -1, respectively, and checks that the loop invariants hold
initially. To execute the loop in a giant step, the variables \w{r}, \w{y}, and
\w{z} are set to the values 0, 0, and 1 from the counterexample, which also
satisfy the loop invariants. The loop condition becomes false, and the
giant-step execution leaves the loop. The execution then fails because the
current value 0 of variable \w{r} contradicts the postcondition.  Since standard
RAC terminated normally but giant-step RAC failed, the proof failure is
categorised as a subcontract-weakness.

See the research report~\cite{becker21rr} for more examples and experimental
results.

\section{Related Work and Perspectives}
\label{sec:concl}

Christakis et al.~\cite{christakis16tacas} use Dynamic Symbolic Execution (DSE)
(i.e., concolic testing) to generate test cases for the part of the code that
underlies a failing VC. The code is extended by run-time checks and fed to the
DSE engine Delfy, that explores all possible paths up to a given
bound. The output can be one of the following: the engine verifies the
method, indicating that the VC is valid and thus no further action is required;
it generates a test case that leads to an assertion violation, indicating that
the VC is invalid; or no test case can be generated in the given bound, where
nothing can be concluded.
Our approach is more directly based on the work of
Petiot et al.~\cite{petiot18fac}. Their work also relies on a DSE engine
(the PathCrawler tool) to classify proof failures and to generate
counterexamples. Each proof failure is classified as
\emph{non-compliance}, \emph{subcontract weakness}, or \emph{prover
  incapacity}. By using DSE, every generated counterexample leads to
an assertion failure during concrete execution of the program.

What distinguishes our approach from the two above is that we derive the test
case leading to a proof failure from the model generated by the SMT solver,
rather than relying on a separate tool such as the DSE engine. Instead of
applying different program transformations, we compare the results of two
different types executions (standard RAC and giant-step RAC) of the original program.

There are quite a lot of potential improvements and extensions to our work,
which are discussed in our research report~\cite{becker21rr}. First, our
approach should be extended to support more features such as the maintenance of
type invariants. Our implementation deserves to be made more robust, to deal
with missing values in the counterexample, and calls to functions that lack an
implementation. A representative example of remaining technical issues is when
the original code makes use of polymorphic data types: in that case, a complex
encoding is applied to the VCs before sending them to provers, and unwinding
this encoding to a obtain an oracle for running the RAC does not currently
work. As seen in the experiments so far, our method is often limited in the RAC,
for example when checking assertions that are quantified formulas, which is a
known issue for RAC~\cite{kosmatov16isola}.
As mentioned in the experiments, our implementation aims to support various
Why3 front-ends, such the one for Ada/SPARK~\cite{mccormick15}. The current
experimental results are not fully satisfying and there are numerous required
practical improvement. Yet, the ability to filter out obviously wrong
counterexamples, and categorise them, is hopefully a major expected improvement
for the SPARK user experience. A remaining challenge is that a counterexample at
Why3's level might not be suitable at the front-end level. In particular,
performing the small-step assertion checking in the
front-end language may result in a more accurate diagnose.

\paragraph{Acknowledgements.} We thank our partners from AdaCore (Yannick Moy)
and Mitsubishi Electric R\&D Centre Europe (David Mentr\'e, Denis Cousineau,
Florian Faissole and Beno\^it Boyer) for their feedback on first
experimentations of our counterexample categorisation approach.

\bibliographystyle{eptcs}
\bibliography{fide}
\end{document}